\documentclass[twocolumn,showpacs,preprintnumbers,amsmath,amssymb]{revtex4}
\usepackage{graphicx}% Include figure files
\usepackage{dcolumn}% Align table columns on decimal point
\usepackage{bm}% bold math
\begin{document}
\newcommand{\Del}{$\Delta$}
\newcommand{\g}{{\rm g}}
\long\def\Omit#1{}
\long\def\omit#1{\small #1}
\def\beq{\begin{equation}}
\def\eeq{\end{equation} }
\def\bea{\begin{eqnarray}}
\def\eea{\end{eqnarray}}
\def\eqref#1{Eq.~(\ref{eq:#1})}
\def\eqlab#1{\label{eq:#1}}
\def\figref#1{Fig.~(\ref{fig:#1})}
\def\figlab#1{\label{fig:#1}}
\def\tabref#1{Table \ref{tab:#1}}
\def\tablab#1{\label{tab:#1}}
\def\secref#1{Section~\ref{sec:#1}}
\def\seclab#1{\label{sec:#1}}
\def\VYP#1#2#3{{\bf #1}, #3 (#2)}  % Volume, page (Year)
\def\NP#1#2#3{Nucl.~Phys.~\VYP{#1}{#2}{#3}}
\def\NPA#1#2#3{Nucl.~Phys.~A~\VYP{#1}{#2}{#3}}
\def\NPB#1#2#3{Nucl.~Phys.~B~\VYP{#1}{#2}{#3}}
\def\PL#1#2#3{Phys.~Lett.~\VYP{#1}{#2}{#3}}
\def\PLB#1#2#3{Phys.~Lett.~B~\VYP{#1}{#2}{#3}}
\def\PR#1#2#3{Phys.~Rev.~\VYP{#1}{#2}{#3}}
\def\PRC#1#2#3{Phys.~Rev.~C~\VYP{#1}{#2}{#3}}
\def\PRD#1#2#3{Phys.~Rev.~D~\VYP{#1}{#2}{#3}}
\def\PRL#1#2#3{Phys.~Rev.~Lett.~\VYP{#1}{#2}{#3}}
\def\AP#1#2#3{Ann.~of Phys.~\VYP{#1}{#2}{#3}}
\def\ZP#1#2#3{Z.\ Phys.\  \VYP{#1}{#2}{#3}}
\newcommand{\etal}{\mbox{\textit et al.}}                       %
\newcommand{\thalf}{\mbox{\small{$\frac{3}{2}$}} }
\def\half{\mbox{\small{$\frac{1}{2}$}}}
\def\quarter{\mbox{\small{$\frac{1}{4}$}}}
\def\sla#1{#1 \hspace{-1ex} \slash}
\def\Sla#1{#1 \hspace{-1.5ex} \slash \hspace{0.5ex} }
\def\eps{\varepsilon}    % photon polarization
\newcommand{\vslash}[1]{#1 \hspace{-0.5 em} /}

\def\olaf{\marginpar{Mod-Olaf}}
\def\her{\marginpar{$\Longleftarrow$}}
\def\bel{\marginpar{$\Downarrow$}}
\def\abo{\marginpar{$\Uparrow$}}

\preprint{KVI/2gam}
\title{Virtual-pion and two-photon production in pp scattering}
\author{O.\ Scholten}
\email{scholten@kvi.nl}
\homepage{http://www.kvi.nl/~scholten}
\affiliation{Kernfysisch Versneller Instituut, University of Groningen, 
9747 AA Groningen, The~Netherlands}
\author{A.\ Yu.\ Korchin\footnote{Permanent address: National Science
Center `Kharkov Institute of Physics and Technology', 61108 Kharkov,
Ukraine} }
\affiliation{Department of Subatomic and Radiation Physics,
University of Gent, B-9000 Gent, Belgium}
\date{\today}

\begin{abstract}
Two-photon production in pp 
scattering is proposed as a means of studying virtual-pion emission. 
Such a process is complementary to  real-pion emission in pp scattering. 
The virtual-pion signal is embedded in a 
background of double-photon bremsstrahlung. We have developed a model to 
describe this background process and show that in certain parts of phase 
space the virtual-pion signal gives significant contribution. In 
addition, through interference with the two-photon bremsstrahlung 
background, one can determine the relative phase of the virtual-pion 
process.
\end{abstract}

\maketitle

\section{Introduction}

Near-threshold pion production in proton-proton scattering has a long
history~\cite{Sta58,Hac78}. More recently it has attracted much
attention after precise data have become available from experiments at
IUCF~\cite{Mey92} and \cite{Bon95}. These showed that this relatively
simple reaction is apparently poorly understood. Earlier works showed
large discrepancy between the calculations and the
data~\cite{Mil91,Nis92}. Later several different mechanisms such as
heavy meson exchange~\cite{Lee93,Hor94}, off-shell structure of the
T-matrix~\cite{Her95O,Ada97}, heavy meson exchange
currents~\cite{VKo96}, and approximations used in the calculation of the
loop contributions~\cite{Sat97} have been proposed to explain this
problem in the earlier calculations. Calculations in a relativistic
one-boson-exchange model~\cite{Eng96} and non-relativistic potential
model~\cite{Han98b} on the other hand appear to reproduce the data
rather well.

In this work we propose to extend the available kinematical regime for
neutral-pion production by investigating the process of virtual 
$\pi^0$ emission which can be observed through its two-photon decay. 
The interest in this process is manyfold. For example,
the importance of off-shell form factors or the off-shell structure of the 
T-matrix~\cite{Her95O,Ada97} can be investigated when an extended 
kinematical regime is available for measuring this reaction. In addition,
studying virtual-pion production below the threshold for real-pion
production in proton-proton scattering implies an important simplification in
the description of the process since the inelasticities, 
always present in the pion-nucleon scattering,
are absent for virtual pions. 
More importantly, interference of two-photon
production via a virtual pion with the background due to two-photon
bremsstrahlung will determine the relative sign of the matrix elements. 
This will allow for a 
better insight in the underlying pion-production process. 
The sign is relevant regarding the discussion of the constructive 
v.s.\ destructive interference of the higher-order diagrams in a 
field-theoretical approach~\cite{Par96,Coh96,Ged99a,Sat97}.

To describe the ``background'', two-photon bremsstrahlung, cross section
we have developed a Soft-Photon Model (SPM) for two-photon emission.

 In this context SPM implies a covariant model satisfying gauge invariance,
which obeys the proper low-energy theorem for small photon momenta
(for example, for the one-photon bremsstrahlung the leading two orders
in powers of the photon energy satisfy model-independent
constraints~\cite{Low58}). As a first step towards such an SPM,
we develop in section II a new SPM for the single-photon bremsstrahlung
amplitude.
This novel SPM, based on a power-series expansion of the T-matrix,
combines ideas of the two SPM's which are frequently used for single-photon
bremsstrahlung in $pp$ scattering: the original SPM~\cite{Nym68,Fea86},
which is directly inspired by the derivation of the low-energy theorem
for bremsstrahlung by Low~\cite{Low58}, and a later one proposed in
ref.~\cite{Lio93}, which has been very successful in reproducing the
observed cross sections~\cite{Lio95,Hui99}.
 The important distinction of
the new SPM from the existing two is that no explicit contact terms, or
the so-called internal contributions, need to be introduced. This
feature makes it the most suitable model for developing the
Two-Soft-Photon Model (2SPM) as discussed at the end of section II. Such
a 2SPM may also be used to calculate the background two-photon signal in
the search for di-baryon states~\cite{Khr01}.

In this work, where the emphasis is placed on the feasibility of
detecting the virtual-$\pi^0$ signal, we have also employed a relatively
simple covariant model to describe the pion-emission process. This
model is discussed in detail in section III. The predictions of this
model are shown to reproduce data on real-pion emission.

In section IV explicit calculations are presented for two-photon 
production where both mechanisms, bremsstrahlung and virtual-$\pi^0$ 
emission, are taken into account. The parts of phase space are indicated 
where the second mechanism is relatively large. It is also demonstrated 
that interference between the two processes is very important.

\section{The soft-photon model}

A starting point in an SPM description of bremsstrahlung in $pp$
scattering is that the dominant -pole- contribution to the amplitude is
derived from the Feynman diagrams where the photon is radiated off the
external legs~\cite{Low58}. To this leading order, several higher-order,
non-pole, terms need to be added which may correspond to meson-exchange,
form-factor, and rescattering contributions. The observation made by Low,
which is the essence of the low-energy theorem~\cite{Low58}, is that in
any description for the amplitude \Omit{(where an expansion is made of
the amplitude in powers of photon energy)} which has the correct pole
structure and is gauge invariant, in a power expansion of the amplitude,
the leading two powers are model independently given by an expression
involving only on-shell (i.e.\ measurable) quantities,
such as the non-radiative NN T-matrix, charge and magnetic moment of
the nucleon.

In the formulation of a SPM description for bremsstrahlung this model 
independence of the leading contributions to the amplitude is 
exploited. In principle, the T-matrix entering in each of the pole diagrams 
needs to be evaluated at different {\it off-shell} kinematics. 
In an SPM one relates the off-shell T-matrix to
the T-matrix at an appropriately chosen on-shell kinematical point.
Based on the low-energy theorem  one can show that effects due to the off-shell
structure of the T-matrix indeed can be ignored to a large extent.
A necessary condition, that the full matrix element for the process is
gauge invariant, is ensured by adding contact terms which are
regular in the limit of vanishing photon momentum.
 As a result one obtains rather
accurate predictions from such a model in spite of its simplicity. Due
to the fact that the SPM's satisfy the low-energy theorem, predictions are
accurate as long as the nucleon-nucleon scattering amplitude varies 
little over an energy range of the order of the photon energy. In the 
past several SPM's have been developed for $pp$ bremsstrahlung. The
earliest one is due to Low and Nyman~\cite{Low58,Nym68} and is based on 
a kind of power series expansion for the amplitude. This particular SPM~\cite{Low58,Nym68,Kor96} 
will be referred to as Low-SPM hereafter. More recently SPM's
were developed by  Liou, Lin and Gibson~\cite{Lio93}, based 
on the explicit evaluation of the tree-level diagrams. The
differences between the different versions lie in the particular 
 choice of the on-shell kinematics at which the T-matrix is
evaluated. Of particular interest for the discussion in this section is
the SPM where the $t$ and $u$ Mandelstam variables are selected to 
define the on-shell kinematics for the T-matrix~\cite{Lio93,Lio95,Kor96},
which will be referred to as the tu-SPM hereafter.

The {\it novel} one- and two-photon bremsstrahlung SPM 
amplitudes developed in the following are
based on a first-order power-series  expansion of the T-matrix around an
appropriately chosen kinematical point (inspired by the Low-SPM) which
is used in the tu-SPM. This hybrid formulation (called here
 pse-SPM) can be shown to correspond to the Low-SPM with the exception of
some additional terms proportional to the magnetic moment multiplied by
derivatives of the T-matrix. On the other hand it would correspond to the
tu-SPM if the T-matrix would depend linearly on the kinematical
variables. The important advantage of our formulation is that the sum of
the diagrams corresponding to radiation off external legs is already
gauge invariant and one therefore does not have to introduce contact
terms.

\subsection{The single-photon bremsstrahlung matrix element}

The antisymmetrized on-mass-shell T-matrix for proton-proton scattering 
can be decomposed in Lorentz scalars~\cite{Gol60,McN83} as
\beq
T(t,u)=\sum_{j=1}^5 C_j(t,u) \Omega_j^{(1)} \cdot \Omega^{j(2)} \;,
\eqlab{T}
\end{equation}
where covariants $\Omega_j$ are taken from the set~\cite{Gol60}
\beq
\Omega_j=\left\{1,\gamma_\mu, \sigma_{\mu \nu}, \gamma_5,
          \gamma_5 \gamma_\mu \right\} \;.
\eqlab{inv1}
\end{equation}
Since only two of the three Mandelstam variables are independent for on-shell 
kinematics ($s+t+u=4m_p^2$) we indicate only the 
dependence on $(t,u)$ of the $T$-matrix and the invariant coefficients $C_j (t,u)$.
In order to arrive at the particular SPM 
(to be referred to as ``pse-SPM"),
which we will later extend to two-photon bremsstrahlung, it is essential
to make a power-series expansion of the coefficients $C_j$ of the T-matrix
around a point ($\bar{t},\bar{u}$) which corresponds to some average
kinematics, 
\bea
C^{pse}_j(t,u)&=&C_j(\bar{t},\bar{u}) +
(t-\bar{t}) \left. {\partial C_j(t,u) \over \partial t}
\right|_{\bar{t},\bar{u}} \nonumber \\ &+&
(u-\bar{u}) \left. {\partial C_j(t,u) \over \partial u}
\right|_{\bar{t},\bar{u}} \;,
\eqlab{C-tay}
\eea
where derivatives are evaluated on-shell~\cite{Kor96}. In order to
guarantee antisymmetry of the bremsstrahlung amplitude under the interchange
of identical particles~\cite{Kor96}, the point ($\bar{t},\bar{u}$) is
defined according to 
\Omit{Alex: this states clearly what has actually been done}
\bea
\bar{s}&=&(s_i+s_f)/2-k^2/6 \;, \nonumber \\
\bar{t}&=&(t_1+t_2)/2-k^2/6 \;, \eqlab{ave-kin} \\
\bar{u}&=&(u_{1'2}+u_{12'})/2-k^2/6 \;, \nonumber
\eea
where $k^2$ is the invariant mass of the emitted particle
($k^2=0$ presently for a single real photon). The Mandelstam variables
are defined as $s_i=q_{s_i}^2$ and similarly for the others, where
 the 4-momenta
\bea
q_{u_{1'2}}=p'_1 - p_2 \; \; &,& \; \; q_{u_{12'}}=p_1 - p'_2 \;, \nonumber \\
q_{t_1}=p_1 - p'_1 \; \; &,& \; \; q_{t_2}=p'_2 - p_2 \;, \eqlab{q} \\
q_{s_i}=p_1 + p_2 \; \; &,& \; \; q_{s_f}=p'_1 + p'_2 \; ,  \nonumber
\eea
are given explicitly in terms of the momenta of the incoming and outgoing
protons for  bremsstrahlung (see also \figref{feyn-1g}).

\begin{figure}[!htb]
\includegraphics[width=8cm,bb=0 460 510  580]{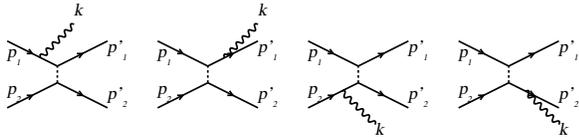}
\caption[pi1]{Feynman diagrams included in the calculation of the 
single-photon bremsstrahlung amplitude. The dotted lines depict the
T-matrix, wavy lines - the photon and solid lines - the proton.
\figlab{feyn-1g}}
\end{figure}

Following ref.~\cite{Lio93} the pole contribution to the amplitude is 
constructed by adding the contributions from the four Feynman diagrams 
corresponding to emission from each of the external legs (see 
\figref{feyn-1g}),
\bea
{\cal M}^\mu&=&\bar{u}_{\lambda'_1}(p'_1) \bar{u}_{\lambda'_2}(p'_2) \; 
\Big[
T_1\;S^{(1)}(p_1-k)\;\Gamma^{\mu(1)}\; \nonumber \\ 
&+& \;\Gamma^{\mu(1)}\;S^{(1)}(p'_1+k)\;T_2 + 
T_3\;S^{(2)}(p_2-k)\;\Gamma^{\mu(2)}\; \nonumber \\
&+& \;\Gamma^{\mu(2)}\;S^{(2)}(p'_2+k)\;T_4
\Big] \; u_{\lambda_2}(p_2) u_{\lambda_1}(p_1)
\eqlab{tay-SPM} \eea
where the index on the T-matrix defines the kinematics at which it is 
evaluated.
For the present SPM (adopted from the tu-SPM) the T-matrix is evaluated at an
on-shell point defined by the same values for the ($t,u$) variables as are
appropriate for the off-shell T-matrix and can be read from the Feynman
diagrams \figref{feyn-1g}. Expressed in terms of the momenta of the in-
and out-going protons these are, 
\bea
T_1=T({u_{1'2}}, {t_2}) \; \; &,& \; \; T_2=T( {u_{12'}}, {t_2}) \nonumber \\
T_3=T( {u_{12'}}, {t_1}) \; \; &,& \; \; T_4=T( {u_{1'2}}, {t_1}) \;.
\eqlab{T1-4}
\eea
For the coefficients $C_j(t,u)$ the power-series expansion \eqref{C-tay}
is used. Note that evaluation of the on-shell T-matrix at $(\bar{u},\bar{t})$
implies for the energy $\bar{s}= 4m_p^2-\bar{t}-\bar{u}$. This value
is different
from the value which could be inferred from diagrams in \figref{feyn-1g}.
The usual expressions for the nucleon  propagator,
$S(p)=i (\sla{p} - m_p +i0)^{-1}$, 
and photon vertex (photon momentum $k$ directed out from the vertex),
\beq
\Gamma^{\mu (j)} =-ie \big[ \gamma^{\mu (j)} - i {\kappa \over 2m_p}
\sigma^{\mu \nu(j)} k_\nu \big] \;,
\end{equation}
have been used, where $j=1,2$ denotes the particle number and $\kappa$ is 
the proton anomalous magnetic moment.

The amplitude given in \eqref{tay-SPM} has the correct pole structure
by construction and, as can be easily checked,
is gauge invariant without having to add contact terms.
Thus the amplitude in \eqref{tay-SPM}  obeys the low-energy
theorem~\cite{Low58} and qualifies as a SPM amplitude.

Comparing the pse-SPM and Low-SPM (the version of ref.~\cite{Kor96})
in some more detail one finds that
most of the terms are identical, with the exception of additional terms 
in  pse-SPM of the type
\bea
&&\kappa \sigma^{\mu\nu(1)} k_\nu {\sla{p}'_1+m_p \over 2 k\cdot p'_1}
\Big[(t_2-\bar{t})
{\partial T \over \partial t} + (u_{12'}-\bar{u}) {\partial T \over \partial
u} \Big]  \nonumber \\
&& -\kappa \Big[(t_2-\bar{t})
{\partial T \over \partial t} + (u_{1'2}-\bar{u}) {\partial T \over \partial
u} \Big] {\sla{p}_1+m_p \over 2 k\cdot p_1} \sigma^{\mu\nu(1)} k_\nu  \nonumber \\
&&+ (1 \leftrightarrow 2) \;.
\eqlab{diff-T,L}
\eea
The obvious notation is used where  ${\partial T \over \partial t}$ 
implies the terms in the Taylor-series expansion of the $T$-matrix that 
contain the derivatives of the coefficients with respect to $t$. It can be shown that
the terms in \eqref{diff-T,L} are of order $k$ and therefore
the difference is beyond the low-energy theorem as one should have expected.

It is also important to mention that the absence of the contact (internal)
contributions in the present SPM is a consequence of the choice $u,t$
as independent variables in the T-matrix. Should $s,t$ (or $s,u$) be
chosen instead, additional contact terms would be required to restore
the gauge invariance (compare {e.g.}, with the original SPM's of
refs.~\cite{Low58,Nym68}).

\subsubsection{Results for single-photon bremsstrahlung}

\begin{figure*}
\includegraphics[height=13cm,bb=90 17 542 700,angle=-90]{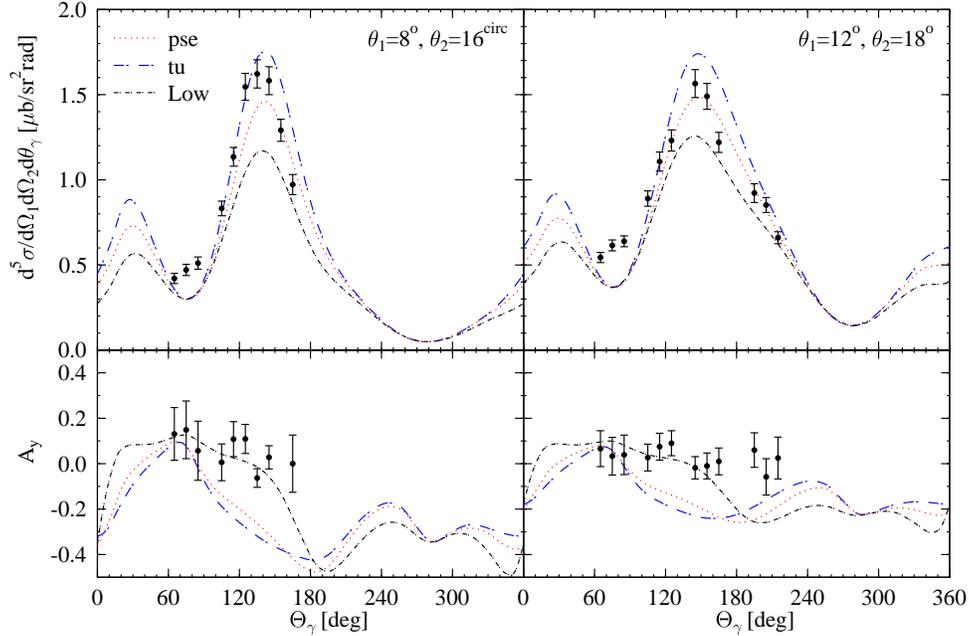}
\caption[pi1]{The differential cross section
$d^5\sigma / d \Omega_{1} d\Omega_{2} d\theta_\gamma$ (upper panel) and
the analyzing power $A_y$ (lower panel) for single-photon bremsstrahlung
as functions of the photon angle.
Prediction of different SPM's (see text) are compared with recent
data obtained at KVI~\cite{Hui99} at a beam energy of 190 MeV.
The angles of the outgoing protons are
kept fixed at $\theta_1=8^\circ$, $\theta_2=16^\circ$ (left panel)
and $\theta_1=12^\circ$, $\theta_2=18^\circ$ (right panel)
for coplanar geometry.  \figlab{t816}}
\end{figure*}

In \figref{t816}  the results for each of the three soft-photon
models are compared with cross-section data obtained in a recent high
precision experiment at KVI at 190 MeV incident energy~\cite{Hui99}. The predictions
of the present pse-SPM appear to lie right in between those of Low-SPM
and tu-SPM. As such the present SPM appears to be in rather good
agreement with the data even though the photon energy is relatively large
(about 80 MeV).

Also shown is the analyzing power $A_y$ for scattering of the
polarized protons. These results suggest that Low-SPM
gives a better description of the data~\cite{Hui99} at 190 MeV
than pse-SPM and tu-SPM. The latter models give close results.

\subsection{The SPM for two-photon bremsstrahlung}

The pse-SPM developed in the above can readily be extended to the case
of two-photon bremsstrahlung since no explicit contact terms (which
appear in the Low-SPM and in the tu-SPM) were introduced. One therefore 
does not
have to deal with the complication of adding a two-photon contact term
or discuss the modification of the single-photon contact term due to the
presence of the second photon.

\begin{figure}[!htb]
\includegraphics[width=8cm,bb=0 190 510  580]{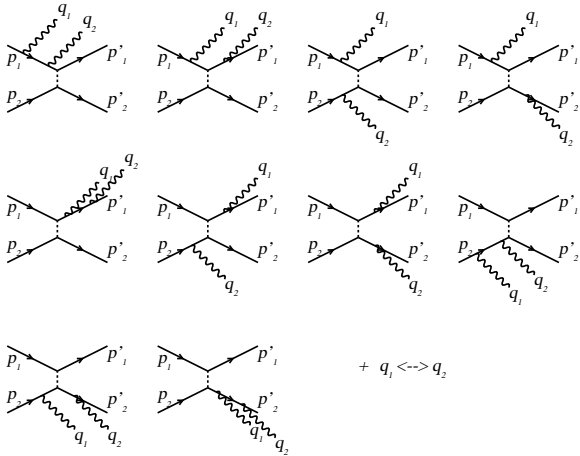}
\caption[pi1]{Feynman diagrams included in the calculation of the  
two-photon bremsstrahlung amplitude. \figlab{feyn-2g}}
\end{figure}

To obtain the two-photon equivalent of pse-SPM (pse-2SPM) one proceeds
in a similar manner as discussed in the previous section. The T-matrix
is written in terms of a power-series expansion around the point of
average kinematics as given in \eqref{ave-kin}, where $k^2$ now equals
 $m_{\gamma \gamma}^2$, the invariant mass
squared of the two-photon system. This particular point is used to preserve
antisymmetry of the matrix element (see ref~\cite{Kor96}). The amplitude can
be constructed by adding the contributions of all diagrams where the two photons
(with momenta $q_1$ and $q_2$, $k=q_1+q_2$) are attached to the
external legs in all possible permutations (see \figref{feyn-2g}), 
\bea
&&{\cal M}(seq)^{\mu\nu}=\bar{u}_{\lambda'_1}(p'_1) \bar{u}_{\lambda'_2}(p'_2)
   \nonumber\\ && \Big[
T_1\;S^{(1)}(p_1-k)\;\Gamma^{\nu(1)}\;S^{(1)}(p_1-q_1)\;\Gamma^{\mu(1)} \nonumber \\ 
&&+\mbox{all possible permutations}
\Big] \; u_{\lambda_2}(p_2) u_{\lambda_1}(p_1)\;,
\eqlab{Tay-S2PM} \eea
where only the first diagram of  \figref{feyn-2g} has been written explicitly.
The $(t,u)$ variables specifying the on-shell point at which the 
T-matrix is evaluated can be easily expressed in terms of the external 
momenta for each diagram.
 However, instead of the true T-matrix the power-series
expansion  \eqref{C-tay} is used. It can
be verified that this amplitude   satisfies gauge invariance,
{\it i.e.} $q_{1 \mu} {\cal M}(seq)^{\mu\nu} =
q_{2 \nu} {\cal M}(seq)^{\mu\nu} =0$,
for the case of radiation off the two-proton system.
The amplitude therefore obeys the low-energy theorem for the two-photon
emission~\cite{Timpr}.

Results for the two-photon bremsstrahlung will be presented in a later section 
together with the results for the virtual-pion amplitude.

\section{Pion production}

 The importance of short-range physics for $\pi^0$ production in
pp scattering was addressed in many references, 
e.g.~\cite{Hor94,Hai96,Ada97,Eng96}.
It was shown that this process is very sensitive to the short-range 
component of the NN interaction which in turns 
reflects in very strong off-shell effects in the 
T-matrix describing the pp rescattering in the  $^1 S_0$ final
state. This partial wave gives the most important contribution
near the pion-production threshold. As a result of the strong
off-shell effects the so-called direct pion production is suppressed,
and other contributions, such as the $\pi N$
rescattering, heavy-meson exchanges, etc.\  
become crucial to obtain agreement with experiment.

In this paper we have opted for simpler, more phenomenological approach 
which still relies on the same
on-mass-shell T-matrix as used in the photon-bremsstrahlung calculations. 
The virtual-pion emission and the sequentional two-photon bremsstrahlung are 
thus described in equivalent models. 
The pion-production amplitude is calculated using radiation off
external legs only while evaluating the T-matrix at a suitably chosen
point corresponding to on-shell kinematics.
We use a general pion-nucleon vertex
\beq
\Gamma_{\pi } (k) =  { G_\pi \over 1+\chi}  \gamma^5
\left( \chi + {\sla{k} \over 2\, m_p} \right)\, ,
\end{equation}
where $\chi$ specifies the admixture of pseudo-scalar (PS) coupling and
$k$ is the pion (outgoing) momentum. The
PS component in the vertex is included to effectively account for
the reaction mechanisms which are not explicitly present in the model.
This issue will be elaborated further on in this  section.

As argued in \cite{Mey92,Han99} the energy dependence of real 
pion production indicates that the final-state interaction 
between the emerging nucleons
should be accounted for correctly. For this reason the T-matrix is
evaluated at an on-shell point corresponding to the same energy, 
$s$, and the same ratio $R=t/u$ as is appropriate for each of 
the four diagrams. 
In should be noticed that for on-shell kinematics the ratio $R$ 
is directly related to the  proton-proton scattering angle. 
% In choosing $(s,R)$ as the important variables,
% the approach for pion production  differs from the one used for 
% bremsstrahlung where we had selected $(t,u)$ instead.
The amplitude which can be read off the diagrams in \figref{feyn-1g} 
(where the photon line is replaced by the pion one) has the form 
\bea
&&{\cal M}_\pi = 
T_1\;S^{(1)}(p_1-k)\; \Gamma_{\pi }^{(1)} (k) + \nonumber\\&& 
\Gamma_{\pi }^{(1)} (k)\;S^{(1)}(p'_1+k) \;T_2 + 
T_3\;S^{(2)}(p_2-k)\; \Gamma_{\pi }^{(2)} (k)\;
\nonumber\\
&&+ \;\Gamma_{\pi }^{(2)} (k)\; S^{(2)}(p'_2+k)\;T_4
\eqlab{pion-ext} 
\eea
with
\bea
T_1=T(s_f, R_1= {t_2}/{u_{1'2}} )\,,\;\;\;
T_2=T(s_i, R_2= {t_2}/{u_{12'}} )\, ,
\nonumber \\
T_3=T(s_f, R_3 = {t_1}/{u_{12'}})\,, \;\;\; 
T_4=T(s_i, R_4=  {t_1}/{u_{1'2}}) \, .  
\eea
The nucleon spinors in \eqref{pion-ext} 
are omitted for brevity.

In the process of calculating the pion-production cross section we
noticed that special care should be paid to the representation of the
T-matrix. Initially the calculation was performed by expanding the T-matrix in
the usual set of five Lorentz covariants given in \eqref{inv1}.

Changing the 
ratio $\chi$ of the PS and pseudo-vector (PV) couplings
by a mere 0.1 would change the real-pion production cross section by about one
order of magnitude. This extreme and unrealistic sensitivity could be
traced back to the unrealistically large coupling to negative-energy
states in the pp system at small energies which is introduced with this
particular choice of covariants.  To avoid the aforementioned
problem we have therefore introduced another set of Lorentz tensors,
chosen such that, when sandwiched between large components of 
positive-energy spinors, they reduce to the five operators usually taken in a
non-relativistic formulation (see for example \cite{Gol64,Ker59,Her95}),
\beq
\Omega^{nr}={1, \vec{\sigma}^{(1)}\cdot\vec{\sigma}^{(2)}, 
i(\vec{\sigma}^{(1)}+\vec{\sigma}^{(2)})\cdot\hat{n}, S_{12}(\hat{t}), 
S_{12}(\hat{u})}\;,
\eqlab{nr-inv}
\end{equation}
where 
$\vec{n}=(\vec{p}_1-\vec{p}_2)\times(\vec{p^\prime}_1-\vec{p^\prime}_2)$, 
$\vec{t}=\vec{q}_t$, $\vec{u}=\vec{q}_u$, the tensor is given by
$S_{12}(\hat{p})=3\; \vec{\sigma}^{(1)}\cdot\hat{p} \; 
\vec{\sigma}^{(2)}\cdot\hat{p}  - 
\vec{\sigma}^{(1)}\cdot\vec{\sigma}^{(2)}$, and the hat denotes a unit vector.
Emphasizing the dependence on $s$ and $R$ of the Lorentz-invariant 
coefficients $C_j$ the  on-shell T-matrix is expressed in a similar 
way as in \eqref{T},
\beq
T(s,R)=\sum_{j=1}^5 C_j(s,R) \Omega_j^{\ (1)} \cdot \Omega^{j(2)} \;.
\eqlab{T-sr}
\end{equation}
A possible choice for the 
covariants $\Omega_j$ is
\beq
\begin{array}{lll}
\Omega_1 = \Sla{Q}_s \;,   & 
\Omega_2 = \gamma_5 \Sla{Q}_n \;, &
\Omega_3 = \gamma_5 \Sla{Q}_p  \;, \\
\Omega_4 = \gamma_5 \Sla{Q}_k\;,  &
\Omega_5 = \Omega_2 + \Omega_1\;, &
\end{array}
\eqlab{inv2}
\end{equation}
defined in terms of the orthogonal four-vectors,
\beq
\begin{array}{l}
Q_s^\mu = (p_1^\mu + p_2^\mu)/W  \; ,\\
Q_k^\mu =  {\cal N}_k (q_t^\mu - (q_t \cdot Q_s) Q_s^\mu) \; , \\
Q_p^\mu =  {\cal N}_p (q_u^\mu - (q_u \cdot Q_s) Q_s^\mu +
(q_u \cdot Q_k) Q_k^\mu)  \; ,\\
Q_n^\mu =  \epsilon^{\mu \nu \sigma \rho}
(Q_{s})_\nu (Q_{k})_\sigma (Q_{p})_\rho \; ,
\end{array}
\end{equation}
normalized such that $Q_s^2 = - Q_k^2 = -Q_p^2 = - Q_n^2 = 1$,
where $W^2=s=(p_1 +p_2 )^2$ and $\ \epsilon^{\mu \nu \sigma \rho}$
is the fully antisymmetric Levi-Civita tensor.
The momenta $q_t$ and $q_u$ are chosen according to \eqref{q}.
It is straightforward to show that in the pp-CM system the
matrix elements of $\Omega_j^{\ (1)} \cdot \Omega^{j(2)}$ for the large
components of the spinors are indeed a linearly independent combination 
of the non-relativistic operators given in \eqref{nr-inv}. The fifth
term in \eqref{inv2}, for example, is the only one that, in the
non-relativistic reduction,  contributes a term like the third one in 
\eqref{nr-inv}. In addition
the matrix elements between large and small components vanish. This set
of five operators is not unique; any linear combination of $q_t$ and $q_u$ 
could have been used to define $Q_k$  and furthermore 
$\Omega_1=1$ is also a valid choice. We have checked that any of these
ambiguities have only minor effects on the calculated pion-production
amplitude.

With the covariants in \eqref{inv2} the cross section for real-pion
 production still depends on the ratio $\chi$ but in a much more
gentle way. 
By varying $\chi$ and keeping a realistic $\pi N$  coupling constant,
the experimentally measured cross section can be reproduced 
with $\chi=1.05$. 
For $\chi=0$, corresponding to the pure PV coupling, the cross 
section is about a factor 10 larger than experiment and appears 
to be independent of the choice of 
covariants, \eqref{inv1} or \eqref{inv2}. The latter feature 
is probably due to the fact that for a PV coupling the 
contribution of negative-energy states is suppressed.

To understand the sensitivity of the cross section to 
the PS component in the $\pi N$ vertex we can  
rewrite the pion-production amplitude in the form     
\bea
{\cal M}_{\pi} = {\cal M}_{\pi}^{(PV)} +{\cal M}_{\pi}^{(cont)}\,,
\eqlab{pion-total}
\eea
where the purely PV contribution is 
\bea
{\cal M}_\pi^{(PV)}&=& i\frac{G_\pi}{2m_p}  
\Big[ T_1\,\gamma_5^{(1)} 
\Big(1-\frac{ m_p \vslash{k}^{(1)}}{ k\cdot p_1 -m_\pi^2/2} \Big)  
\nonumber\\
&& + \Big(1-\frac{m_p \vslash{k}^{(1)}}{k\cdot p'_1 +m_\pi^2/2} \Big)  
\gamma_5^{(1)} \,T_2 
\nonumber \\ 
&& + T_3\, \gamma_5^{(2)}
\Big(1-\frac{m_p \vslash{k}^{(2)}}{k\cdot p_2 -m_\pi^2/2} \Big) 
\nonumber\\
&&+ \Big(1 -\frac{m_p \vslash{k}^{(2)}}{k\cdot p'_2 +m_\pi^2/2} \Big)
\gamma_5^{(2)} \, T_4 
\Big] \,,
\eqlab{pion-PV}
\eea
and  ${\cal M}_{\pi}^{(cont)} $ has a form of a 5-point contact
vertex,
\bea
{\cal M}_{\pi}^{(cont)} &=& -i \Big( \frac{\chi}{1+ \chi } \Big) 
\frac{G_\pi}{2m_p} 
 \Big( T_1\,\gamma_5^{(1)} +\gamma_5^{(1)} \,T_2 
\nonumber\\
&& + T_3\, \gamma_5^{(2)}+\gamma_5^{(2)} \, T_4 \Big) \,.
\eqlab{pion-cont}
\eea

The important observation is that in the soft-pion limit ($k \to 0 $)
the amplitude ${\cal M}_\pi^{(PV)}$ fulfills the requirement of the chiral
symmetry \cite{Nam62} (to be precise, the limit
in \eqref{pion-PV} shoud be taken in the order: $ \lim_{ m_\pi \to 0} 
\{ \lim_{\vec{k} \to 0}... \} $). In this case, of course,
$T_1=T_2=T_3=T_4$ is the T-matrix calculated in kinematics of 
on-shell pp scattering. 
However the pion mass is finite, and the T-matrices in the dominant 
diagrams, $T_1$ and $T_3$,
corresponding to the pion emission off the initial legs, in fact should 
enter far off-shell (for example, the corresponding off-energy-shell
CM momentum is about ${(m_\pi m_p)}^{1/2} \approx 370$
MeV). Due to a strong off-shell dependence of the T-matrix 
(see, e.g. \cite{Hai96,Her95}) a sizable reduction of the cross section
calculated only with ${\cal M}_{\pi}^{(PV)} $ should occur.
The contact term in \eqref{pion-cont} effectively accounts for this 
effect, as well as the other important mechanisms which act in the 
opposite direction, such as (off-shell) $\pi N$ rescattering and 
heavy-meson exchanges, and possible genuine contact vertices in the
underlying Lagrangian.
We will consider $\chi $ as a phenomenological parameter of the model.

Our approach has a certain  similarity to the soft-pion model of 
ref.~\cite{Sch69}, where the authors applied yet another 
method to account for off-shell effects, and found a 
reasonable agreement with the data available at the time.

%The sensitivity of the cross section on $\chi$ in this simple 
%model can be regarded as a 
%reflection of the fact that in a fully microscopic calculation the cross 
%section is small as the 
%result of a rather sensitive cancellation between the one-pion-exchange 
%part in the potential and the $\sigma$- and $\omega$- exchange 
%parts~\cite{Nakpr}. The importance of short-range physics for the 
%pion-production cross section was also addressed in many other 
%references, e.g.~\cite{Hor94,Ada97,Eng96}.
% Another point of view is that instead of introducing the pseudo-scalar 
%structure in the pion vertex we could also have taken an appropriately 
%chosen (5-point) contact term. %\Omit%
%{Such contact terms are generally  
%introduced to restore a (gauge) symmetry, which in this case would be 
% chiral symmetry.}

\begin{figure}
\includegraphics[height=8.5cm,angle=-90,bb=144 19 555 674]{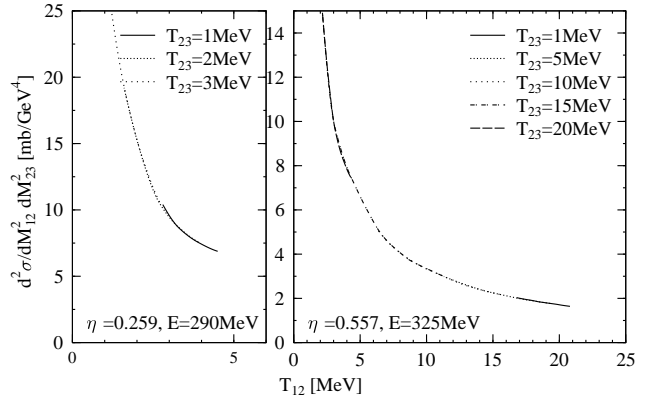}
\caption[pi1]{The angle-integrated cross section for real-pion production 
is plotted versus the relative kinetic energy $T_{12}$ 
of the outgoing two-proton system. Results are shown for two incoming
proton energies and for a selection of $\pi$-N relative kinetic
energies.
\figlab{pi-t12}}
\end{figure}

The angle integrated cross section for real-$\pi^0$ production is plotted in 
\figref{pi-t12} as function of the relative kinetic energy in the final 
two-proton system, defined as $T_{12}=M_{12}-2m_p$,
where $M_{12}$ is the invariant mass of the two-proton system. 
It can be seen that to a large 
extent the cross section is independent of $T_{23}=M_{23}-m_p-m_{\pi}$ 
and falls off roughly proportional to $1/T_{12}$ in accordance with 
\cite{Mey92}.

\begin{figure}
\includegraphics[width=4cm,bb=40 20 387 485]{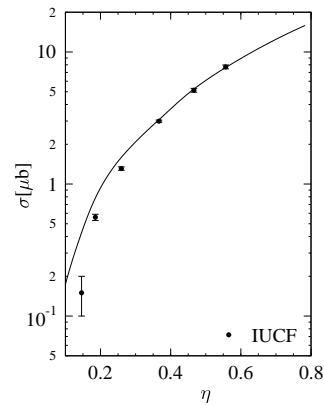}
\caption[pi1]{The integrated cross section for real-pion production as 
function of $\eta$ which is related to the beam energy through 
\eqref{eta}. The data are from~\cite{Mey92}.
\figlab{pi-tot}}
\end{figure}

The total $\pi^0$ production cross section as function of $\eta$ (see 
\eqref{eta} for the definition) is compared to the data of 
\cite{Mey92} in \figref{pi-tot}.
It is seen that the calculation agrees
well with the data  in both magnitude and energy dependence.
As such we 
conclude that this simple model is able to give a reasonable estimate 
of the pion-production cross section and will thus use it also in the
calculation of virtual-pion production discussed in the following 
section.

\subsection{Virtual pions}

The amplitude for two-photon emission mediated by a virtual pion can be
factorized in two terms. The first is the amplitude for virtual-pion
production, ${\cal M}_\pi $,
which is identical in structure to the one for real pions. The second term
describes the decay of the virtual pion. The amplitude in question now reads
\begin{eqnarray}
{\cal M}^{\mu\nu}({\pi})
&=&  \frac{i e^2 g_{\pi \gamma\gamma}}{(k^2 - m^2_\pi +i0 ) m_\pi }
\epsilon^{\mu\nu\alpha\beta}
q_{1\alpha}q_{2\beta} \nonumber \\&&
\times {\cal M}_\pi (p'_1, p'_2; p_1, p_2)\;,
\eqlab{pion1}
\end{eqnarray}
where $k = q_1 + q_2$ is the momentum of the virtual pion and $g_{\pi \gamma\gamma}
\approx 0.0375$ is the $\pi^0 \rightarrow \gamma\gamma$ decay constant.
The total amplitude for the $pp \to pp \gamma \gamma$ process is
\begin{eqnarray}
{\cal M}^{\mu\nu} = {\cal M}^{\mu\nu}(seq) +{\cal M}^{\mu\nu}(\pi) \;,
\eqlab{total_amplitude}
\end{eqnarray}
the sum of the amplitudes given in \eqref{Tay-S2PM} and \eqref{pion1}

\section{Results}

The calculations presented in this section are done for an incoming proton 
energy of 280 MeV in the Lab system which is just below the 
pion-production threshold ($2m_\pi + m_\pi^2 /2 m_p$).

\begin{figure*}
\includegraphics[height=13cm,angle=-90,bb=180 23 550 723]{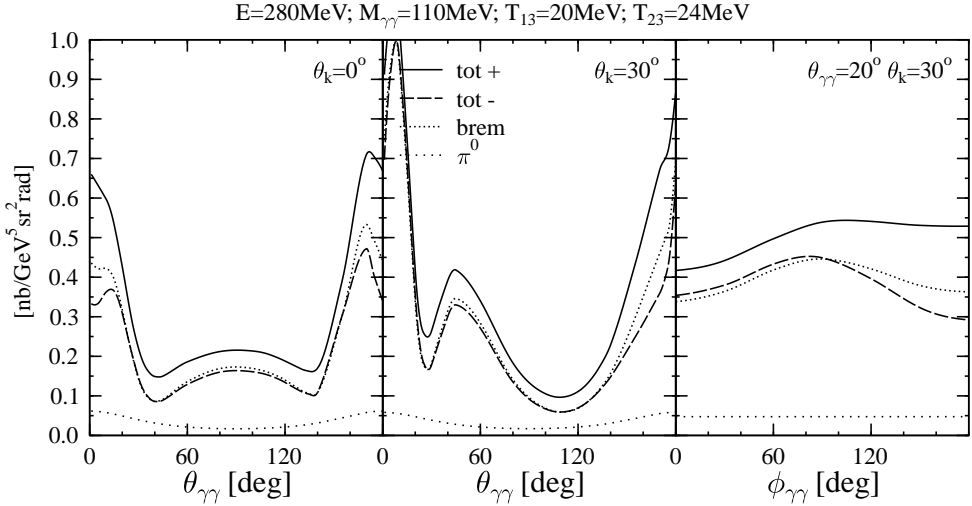}
\caption[pi1]{The dependence of the differential cross section 
$d^8\sigma/dm_{\gamma\gamma}dM^2_{12}dM^2_{23}d\Omega_{\gamma\gamma}d\Omega_kd\phi_1$ 
on the two-photon angles. The two-photon bremsstrahlung 
result is given by the dense-dotted curve, the pure virtual-pion process by the 
sparse-dotted curve, the full result by the drawn curve, while the 
dashed curve gives the coherent sum when changing ad-hoc the sign of
the virtual-pion contribution.\figlab{tetgg}}
\end{figure*}

In \figref{tetgg} the cross section for two-photon production is plotted
for certain exclusive kinematics. We have opted to use the Dalitz
coordinates (see Appendix B for a more detailed discussion) for expressing
the differential cross sections, specified by i) $T_{12}$, the relative
energy between the two protons; ii) $T_{13}$, the relative energy
between a proton and the sum-momentum of the two photons (equal to the
momentum of the virtual pion); iii) the Euler angles of the plane
spanned by the two protons and the virtual pion with respect to the
incoming beam direction, i.e.\ ($\theta_{k}, \phi_{k}$), and where the
third angle is trivial due to azimuthal symmetry. This is supplemented
by the angles ($\theta_{\gamma \gamma}, \phi_{\gamma \gamma}$)
specifying the orientation of the two-photon relative momentum in their
c.m.\ frame and the invariant mass $m_{\gamma\gamma}$ of the two-photon system
or -equivalently- the virtual pion. These coordinates can be used for the
sequential two-photon emission as well as for the virtual-pion process. 
We have used these coordinates instead of the traditionally adopted
ones in bremsstrahlung for a few reasons. Firstly, the phase space
factor is a very smooth (in many cases independent) function of the kinematical
variables and differential cross sections thus directly reflect the magnitude
of the underlying matrix element. Secondly, the absence of divergencies of the
phase space factor allows for a straightforward evaluation of (partially)
integrated cross sections. Thirdly, these coordinates uniquely determine the
kinematics of the event while in polar coordinates a kinematical solution is not
always uniquely defined (this happens in very selected parts of phase space
only). It should be noted that $T_{23}$ is related to $T_{12}$ and $T_{13}$ by a
simple algebraic relation.

In the figures the two-photon cross sections due to the intermediate
virtual-$\pi^0$ mechanism and the sequential two-photon emission are
indicated separately. There is a strong interference between the two
contributions, the total (from adding the amplitudes, labelled 'tot+' in
the figures) is larger than the sum of the individual cross sections. The
importance of interference for the total cross section implies that the
cross section is sensitive to the relative phase between the amplitudes of
the uncorrelated and the virtual-pion two-photon emission processes. To
show this, we also plot the cross section for the case in which the
virtual-pion matrix element has been arbitrarily, only for display purposes, 
multiplied with a minus
sign (i.e.\ changing the relative sign in \eqref{total_amplitude} eventhough the sign given there is correct). 
This calculation, labeled 'tot-' in \figref{tetgg}, gives rise to a much
smaller total cross section. It should be noted that changing the sign
does not affect the real-pion production cross section since it
is independent of the sign. From
\figref{tetgg} it can be seen that the angular distributions depend on the
phase of the virtual-pion contribution, however the largest effect of
changing the sign shows up in the overall magnitude. We have checked that
this is the case in a large region of phase space and the distributions
shown here can be regarded as typical.

Another aspect which can be seen from \figref{tetgg} is that the angular
distributions of the sequentional two-photon emission process show
pronounced structures. This is to be expected as the single-photon
bremsstrahlung angular distribution shows pronounced peaks which
are due to the quadrupole nature of the electric radiation and the interference
with magnetic radiation. The virtual-pion mechanism has a rather featureless
distributions, due to the fact that the two photons couple to the quantum
numbers of the pion, $J^\pi=0^-$.

It is also apparent from \figref{tetgg} that --unfortunately-- one 
cannot point to a particular feature in the angular distribution which 
is especially sensitive to the virtual-pion contribution. There are no 
quantum numbers that distinguish this process from the 
bremsstrahlung contribution.

For the above reason, and also because cross sections are --in general-- 
small for two-photon emission, we
investigate whether the virtual-pion signal can also be seen
in less exclusive kinematics where certain angles have been integrated. 
Since, as remarked before, the virtual pion contribution seems to give 
primarily rise to an overall increase of the cross sections we have 
performed a simple integration of the differential cross section.

\begin{figure}
\includegraphics[height=8cm,angle=-90,bb=6 0 553 514]{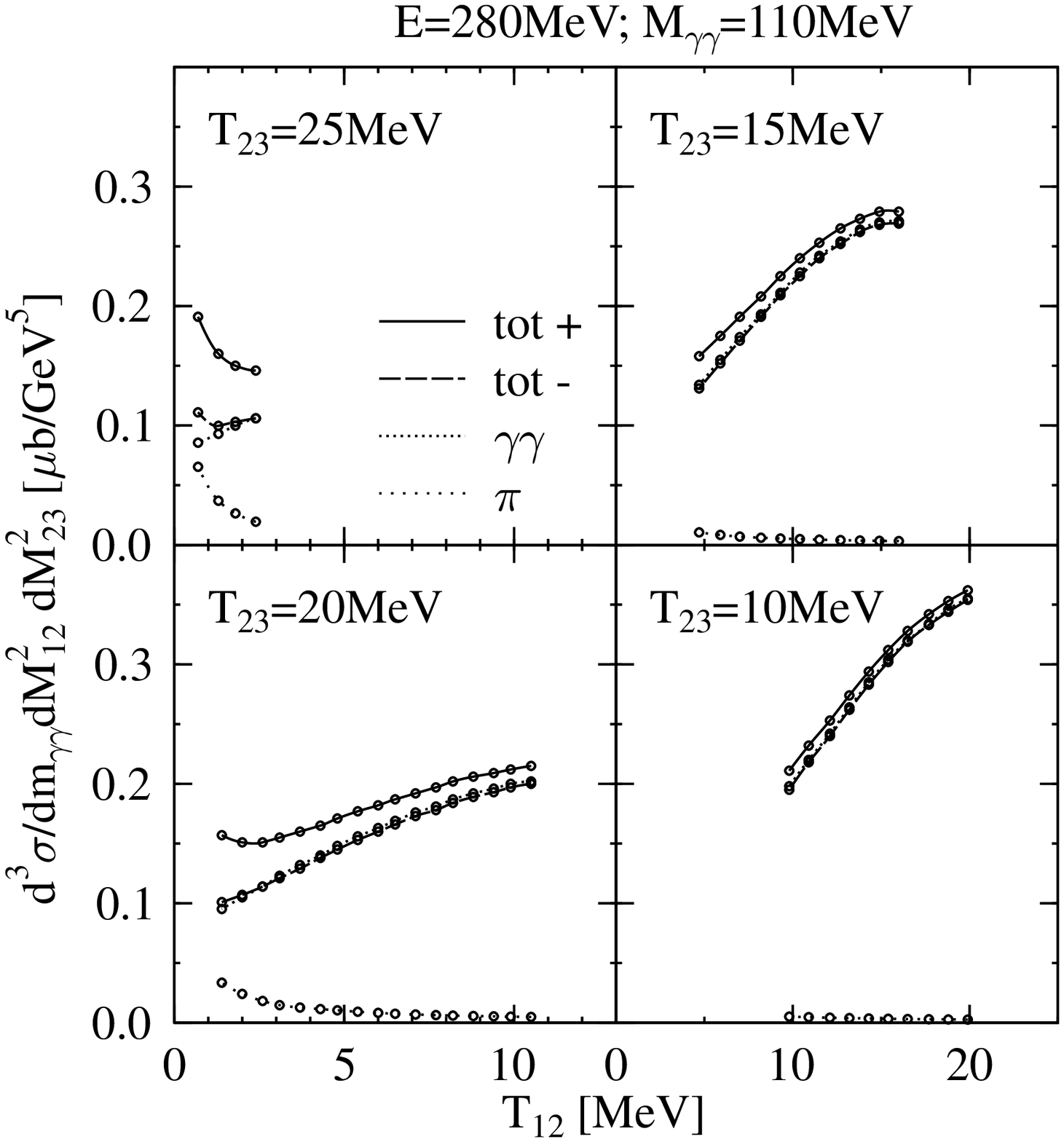}
\caption[pi1]{The cross section for two-photon production, integrated
over angles, is plotted versus the relative kinetic energy $T_{12}$ of
the final two-proton system. Results are shown for a selection of
$\pi$-N relative kinetic energies. The meaning of the curves is the same 
as in \figref{tetgg}. The end points of the curves are determined by kinematics.
\figlab{VG-a}} 
\end{figure}

The squared matrix element for pion emission is inversely proportional to 
the relative energy in the final pp system~\cite{Mey92}. One thus expects that the 
virtual-pion process is most pronounced for the lowest values of 
$T_{12}$. This is indeed supported by our calculations as shown in 
\figref{VG-a}, where the difference between the 
full calculation (labeled 'tot+') and the sequential two-photon 
process strongly depends on $T_{12}$ and hardly on $T_{23}$. In 
\figref{VG-a} all angles have been integrated.

\begin{figure}
\includegraphics[width=6cm,bb=34 20 390 400]{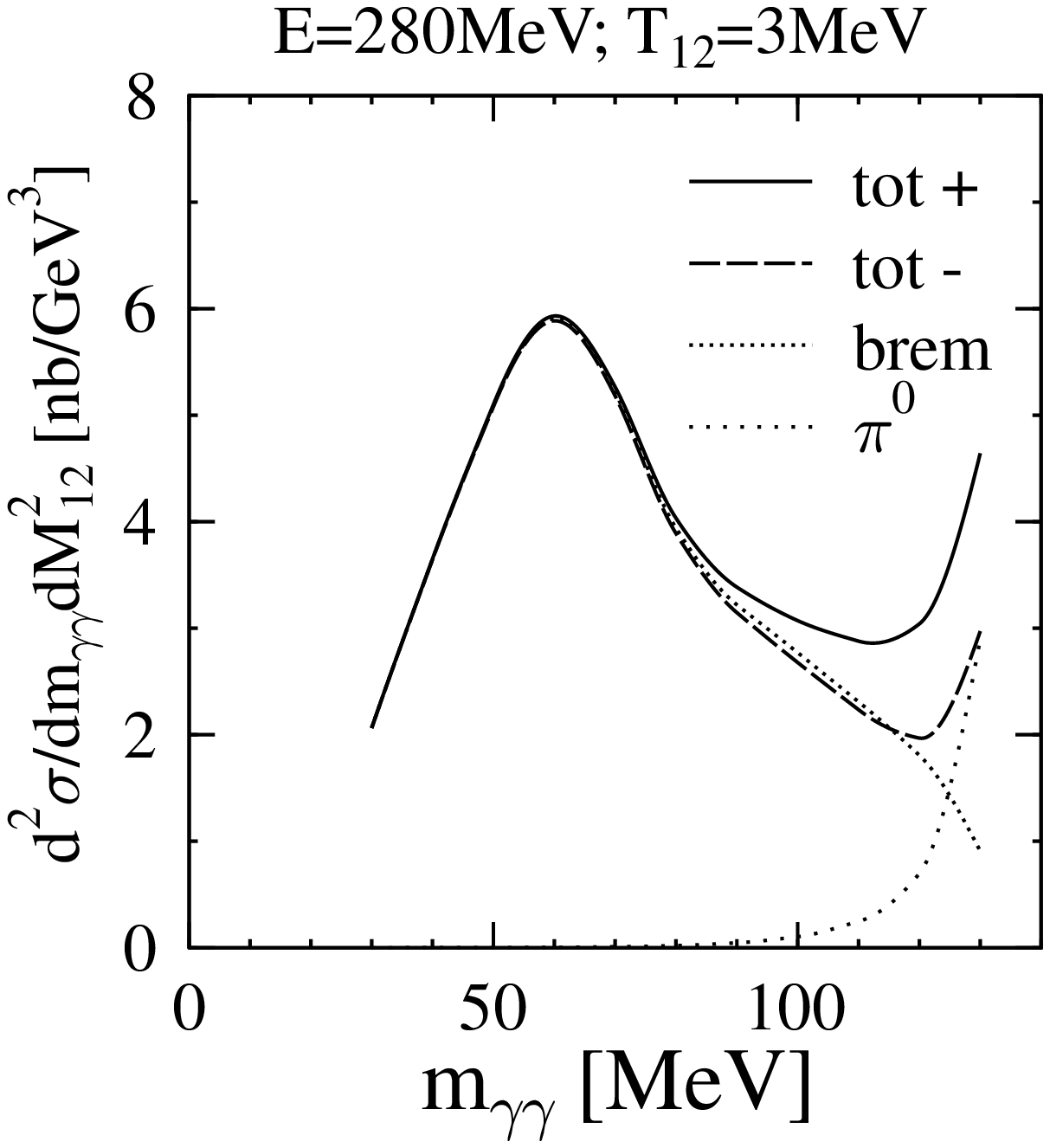}
\caption[pi1]{The cross section for two-photon production, integrated 
over all variables at fixed relative kinetic energy $T_{12}=3$ MeV
of the final two-proton system, is plotted as function of the
two-photon invariant mass. The meaning of the curves is the same 
as in \figref{tetgg}.\figlab{int3}}
\end{figure}

The unambiguous signature of the virtual-pion contribution is that it increases
the closer one approaches the real-pion pole. This can clearly be seen from
\figref{int3} where the cross section is shown as function of 
$m_{\gamma\gamma}$
at fixed $T_{12}$. All other variables, i.e.\ all angles and also $T_{23}$, are integrated. Even this rather 
inclusive cross section shows a clear sensitivity to the interference 
between the sequential and the virtual-pion two-photon emission 
processes.

\section{Summary and conclusions}

In this work we have shown that the two-photon bremsstrahlung offers the
interesting possibility to `measure'  subthreshold pion
production in $pp$ scattering.
It allows for studying pion production in
kinematics which is not accessible in the $pp \rightarrow pp \pi^0$ reaction.
In addition, the phase of the virtual-pion process
with respect to that of sequential two-photon emission can be investigated.

To account for the sequential two-photon emission process,
which is an important background,
a novel soft-photon model (called pse-SPM) is developed.
This model is tested in a calculation of single-photon bremsstrahlung,
and is shown to give accurate results for cross sections.

Calculated exclusive cross sections of the $pp \rightarrow pp \gamma\gamma$
reaction are in general small, however sensitivity of the cross
sections to the
virtual-pion signal remains even for rather inclusive cross sections.

\begin{acknowledgments}
Part of this work was performed as part of the research program
of the Stichting voor Fundamenteel Onderzoek der Materie
(FOM) with financial support from the Nederlandse Organisatie
voor Wetenschappelijk Onderzoek (NWO). One of the authors (A.Yu.K.)
acknowledges a special grant from
the NWO. He would also like to thank the staff
of the Kernfysisch Versneller Instituut in Groningen for the kind hospitality.
We acknowledge discussions with R.~Timmermans.
We thank K.~Nakayama for his help in checking the calculation of 
the pion-emission process and J.~Bacelar for a careful reading of the 
manuscript and discussions on the feasibility of measuring the different 
observables.
\end{acknowledgments}

\appendix

\section{Kinematics for two-photon production}

For the reaction $\ N+N\rightarrow N+N+\gamma +\gamma $ the momenta are
denoted by $\ p_{1}, p_{2}, p_{1}^{\prime}, p_{2}^{\prime }, 
q_{1},q_{2}$  (see \figref{feyn-2g}).
Energy-momentum conservation reads \ $p_{1}+p_{2}=p_{1}^{\prime
}+p_{2}^{\prime }+q_{1}+q_{2}$. The cross section is
\begin{eqnarray*}
d\sigma &=& \frac{m_p^{4}}{j} \int |A|^{2} (2\pi )^{4}
\delta^{4}(p_{1}+p_{2}-p_{1}^{\prime }-p_{2}^{\prime }-q_{1}-q_{2})\\
&&\times \frac{d^{3}p_{1}^{\prime }}{(2\pi )^{3}E_{1}^{\prime }}
\frac{d^{3}p_{2}^{\prime }}{(2\pi )^{3}E_{2}^{\prime }}
\frac{d^{3}q_{1}}{(2\pi )^{3}2\varepsilon_{1}}
\frac{d^{3}q_{2}}{(2\pi )^{3}2\varepsilon_{2}}\,,
\end{eqnarray*}
where $A={\cal M}^{\mu\nu} \epsilon^{*}_{1\mu} \epsilon^{*}_{2\nu}$ is
the invariant amplitude, $\epsilon_1$ and $\epsilon_2$ are the
polarization vectors of the photons, $\varepsilon_1 =|\vec{q}_1|$,
$\varepsilon_2 =|\vec{q}_2|$, and $j=\sqrt{(p_{1}\cdot
p_{2})^{2}-m_p^{4}}=m_p|\vec{p}_{Lab}|$ in the laboratory frame where
$p_{2}=(m_p, \vec{0})$. Using the identity 
$\int \delta^{4}(q_{1}+q_{2}-k)d^{4}k=1$ 
the cross section is put in the form
\begin{eqnarray}
d\sigma &=& \frac{m_p^{4}}{(2\pi )^{8}j} \int |A|^{2}
  \delta ^{4}(p_{1}+p_{2}-p_{1}^{\prime }- p_{2}^{\prime }- k)
  I_{\gamma\gamma } \nonumber \\
&&\times \frac{d^{3}p_{1}^{\prime }}{E_{1}^{\prime }}
  \frac{d^{3}p_{2}^{\prime }}{E_{2}^{\prime }} d^{4}k\,,
\eqlab{sig2}
\end{eqnarray}
where $I_{\gamma\gamma}$ is the two-photon phase-space integral defined as
\begin{eqnarray*}
I_{\gamma\gamma} = \int \delta ^{4}(q_{1}+q_{2}-k)
  \frac{d^{3}q_{1}}{2\varepsilon_{1} }
  \frac{d^{3}q_{2}}{2\varepsilon_{2} }\,.
\end{eqnarray*}
To calculate this integral in an arbitrary frame we introduce the relative 
and total 4-momenta of the photons
\begin{eqnarray}
&&k=q_{1}+q_{2},\;\;\;\;\;\; l=\frac{1}{2}(q_{1}-q_{2}),\nonumber \\
&&q_{1}=\frac{1}{2}k+l, \;\;\;\;\;\; q_{2}=\frac{1}{2}k-l\,.
\end{eqnarray}
The Jacobian of the transformation from $\vec{q}_1,\ \vec{q}_2$ to
$\vec{l},\ \vec{k}$ is unity, and after removing the trivial
$\delta$-function we get
\begin{eqnarray}
I_{\gamma\gamma } &=& \int \delta(\varepsilon_{\frac{1}{2}\vec{k}+\vec{l}}
 +\varepsilon_{\frac{1}{2}\vec{k}-\vec{l}}-k_{0})
  \frac{d^{3}l}{2\varepsilon_{\frac{1}{2}\vec{k}+\vec{l}} \,
                2\varepsilon_{\frac{1}{2}\vec{k}-\vec{l}}}\,
\nonumber \\
&=&\frac{|\vec{l}|^{2}}{4(|\vec{l}|k_{0}-|\vec{k}|
l_{0}\cos \theta _{\gamma\gamma})}d\Omega_{\gamma\gamma}\,,
\eqlab{igg2}
\end{eqnarray}
with $d\Omega_{\gamma\gamma}= \sin \theta_{\gamma\gamma}
 d\theta_{\gamma\gamma} d\phi_{\gamma\gamma}$,
 where we introduced the polar and azimuthal angles 
$\theta_{\gamma\gamma}$ and $\phi_{\gamma\gamma}$ between the 3-vectors $\vec{k}$
and  $\vec{l} $. For real photons ($q_{1}^{2}=q_{2}^{2}=0$) one can show 
that $ k\cdot l=0$ and $4\,l^{2}+m_{\gamma\gamma}^{2}=0$, 
where $m_{\gamma\gamma }^{2} = k^{2}$ is the invariant mass of the
two-photon system. Expressing now $l_{0}$ in terms of the 3-momentum
$|\vec{l}|$  we obtain
\begin{eqnarray*}
|\vec{l}| = \frac{m_{\gamma\gamma }}{2\sqrt{1-\frac{\vec{k}^{2}}{k_{0}^{2}}
\cos^{2}\theta_{\gamma\gamma}}}
= \frac{m_{\gamma\gamma }k_{0}} {2\sqrt{m_{\gamma\gamma }^{2}+ 
\vec{k}^{2}\sin^{2}\theta_{\gamma\gamma}}},
\end{eqnarray*}
with $k_{0}=\sqrt{m_{\gamma\gamma }^{2}+\vec{k}^{2}}.$
The two-photon phase space \eqref{igg2} can be simplified to
\begin{eqnarray}
I_{\gamma\gamma} = \frac{|\vec{l}|^{3}}{k_{0}m_{\gamma\gamma }^{2}}
   d\Omega_{\gamma\gamma}\,.
\end{eqnarray}

As a last step the integration over $k_{0}$ in \eqref{sig2} is replaced by
an integration over the two-photon invariant mass using
$k_{0} dk_{0}= m_{\gamma\gamma }dm_{\gamma\gamma}$. We obtain
\begin{eqnarray}
d\sigma =\frac{2 m_p^{4}}{(2\pi )^{8}j}
\int |A|^{2} J(m_{\gamma\gamma }) I_{\gamma\gamma }
 m_{\gamma\gamma } dm_{\gamma\gamma }\,,
\end{eqnarray}
where we introduced the 3-particle phase-space integral
\bea
J(m_{\gamma\gamma })&=& \int
\delta^{4}( p_1 + p_2 - p_1^\prime - p_2^\prime - k)\nonumber \\
&&\times \frac{ d^3 p_{1}^{\prime }}{E_{1}^{\prime }}
\frac{d^3 p_{2}^{\prime }}{E_{2}^{\prime }}
\frac{d^3 k}{2 k_{0}}\,.
\eqlab{Jac}
\eea
In the one-photon bremsstrahlung the similar integral is traditionally evaluated
in polar coordinates (see, for example, \cite{Kor96}) leading to the
cross section of the type shown in \figref{t816}.
For the two-photon bremsstrahlung in the present paper we will use the Dalitz
coordinates instead, as discussed in Appendix B.

\section{Dalitz coordinates}

To evaluate the phase-space integral in \eqref{Jac} we choose the CM frame where
$\vec{p}_{1} +\vec{p}_{2} =0$, and carry out the
integration over  $\vec{p}_2^{\,\prime}$.
Introducing $s = (p_{1}+p_{2})^{2}$ we obtain
\[
J(m_{\gamma\gamma}) =\int \delta( \sqrt{s}-E_1^{\prime} -E_{2}^{\prime}- k_0 )
\frac{d^3 p_1^{\prime}}{E_1^{\prime}E_2^{\prime}} \frac{d^3 k}{2 k_0 }\,,
\]
where  $E_2^{\prime2}= m_p^2+ \vec{p}_2^{\,\prime2} =
m_p^{2}+(\vec{p}_1^{\,\prime}+\vec{k})^{2}
=m_p^{2}+\vec{p}_1^{\,\prime2} +\vec{k}^{2}+2 p_{1}^{\prime} k\cos
\theta_{13}$.  Using
\begin{eqnarray*}
&&d^{3} p_{1}^{\prime} \, d^{3}k
=p_1^\prime \, E_1^\prime \, dE_1^\prime \, d\Omega_1\,
k \, k_0 \, dk_0 \, d\Omega _{k} \\
&&= p_1^\prime \, E_1^\prime \,dE_1^\prime\, d\cos \theta_{13} \, d\phi_1 \,
k \, k_0\, dk_0\, d\cos \theta_k\, d\phi_k
\end{eqnarray*}
and integrating over $\cos \theta_{13}$ using the $\delta$-function we 
obtain
\[
J(m_{\gamma\gamma})={1 \over 2} d\cos \theta_{k}\,d\phi_k\, d\phi_1\,
dk_0 \, dE_1^{\prime} \;.
\]
Defining invariant masses
\begin{eqnarray}
M_{12}^{2} &=& (T_{12}+2m_p)^2 = (p_{1}^{\prime}+p_{2}^{\prime})^{2}
\nonumber \\
&=& s-2\sqrt{s} k_0 +m_{\gamma\gamma}^{2} \;, \nonumber \\
M_{23}^{2} &=& (T_{23}+m_p+m_\pi)^2 = (p_2^\prime+k)^2 \nonumber \\
&=&s-2\sqrt{s} E_1^\prime + m_p^2,
\eqlab{inv-mass}
\end{eqnarray}
we can cast the integral $J(m_{\gamma \gamma})$ in the form
\beq
J(m_{\gamma\gamma}) = {1 \over 8s} d\cos \theta_{k}\, d\phi_{k}\, d\phi_1\,
d\,M_{12}^{2} dM_{23}^{2}\,.
\eqlab{D-jac}
\end{equation}
Here  the angles $\phi _{k},\,\theta _{k},\,\phi_1$\ describe in the CM frame the
orientation of the plane in which the momenta lie of the 
outgoing two protons and the two-photon system (the virtual pion) 
with respect to the incoming beam. Specifically, the 
 angle $\phi_{1}$ is defined as the azimuthal angle of the momentum 
$\vec{p}_1^{\,\prime}$
in the frame, where momentum $\vec{k}$ is along OZ axis and the OX axis 
lies in the plane spanned by the beam and $\vec{k}$. The angle
$\theta_{k}$ is taken as the angle in the CM frame between the incoming 
momentum and $\vec{k}$ while $\phi_k$ is a trivial azimuthal angle. 
In the CM frame the momenta can now be expressed as
\begin{eqnarray*}
&&\vec{k} =(k_x,\; k_y,\; k_z) = k(\sin \theta_{k},\; 0,\; \cos \theta_{k}),\, \\
&&\vec{p}_1^{\,\prime} = (p_{1x}^{\prime},\; p_{1y}^{\prime},\; 
p_{1z}^{\prime})= \\
&&=p_{1}^{\prime}( \cos\theta_{k} \sin\theta_{13} \cos\phi_{1}
+ \sin\theta_{k} \cos\theta_{13},\\
&&\;\sin \theta_{13} \sin\phi_{1}, \;
- \sin\theta_{k} \sin\theta_{13} \cos\phi_{1}+ \cos\theta_{k} \cos\theta_{13}), \\
&&\vec{p}_2^{\,\prime} = -\vec{p}_1^{\,\prime} -\vec{k}.
\end{eqnarray*}
The  magnitudes of  $\vec{k}$, $\vec{p}_1^{\,\prime}$ and 
$\vec{p}_2^{\,\prime}$ are determined
through the energies of the two-photon system and nucleons
\begin{eqnarray*}
k_0 &=&\frac{s-M_{12}^{2}+m_{\gamma\gamma}^{2}}{2\sqrt{s}}\;, \\
E_{1}^{\prime}&=&\frac{s-M_{23}^{2}+m_p^{2}}{2\sqrt{s}}\; ,\\
E_{2}^{\prime}&=&\sqrt{s}-E_{1}^{\prime}- k_0 \,,
\end{eqnarray*}
and the angle between $\vec{p}_1^{\,\prime}$  and $\vec{k}$ can be expressed 
as,
\[
\cos \theta_{13}=( E_{2}^{\prime2}-E_{1}^{\prime2}- k_0^{2} +
m_{\gamma\gamma}^{2} )/ {2kp_{1}^{\prime}}.
\]

So far the momenta were defined in the CM frame. The boost to the Lab system is specified by the velocity
$V= {p_{1}}/ ({m_p+E_1}) $  and the Lorentz-factor
$\gamma = ({m_p+E_1}) /{\sqrt{s}}$.
 The Z components of the vectors in the lab can now be expressed as
\begin{eqnarray*}
k_{z}^{Lab } &=&\gamma (k_{z}+V k_0 ), \\
p_{1z}^{\prime Lab } &=& \gamma (p_{1z}^{\prime} +V E_{1}^{\prime}), \\
p_{2z}^{\prime Lab } &=&p_{1}-k_{z}^{Lab }-p_{1z}^{\prime Lab },
\end{eqnarray*}
while the X  and Y components do not change.

\section{Total cross section for pion production}

Using the Dalitz coordinates, the cross section for real pion 
production is written as
\begin{equation}
d\sigma =\frac{m_p^{4}}{(2\pi )^{5}j}|{\cal M}_\pi |^{2} J(m_\pi)
\eqlab{CS-pi}
\end{equation}
where particle 3 is associated with the pion and $J(m_\pi)$ is given
in \eqref{D-jac} where $m_{\gamma\gamma} \to m_\pi$.

Let us first make a tentative assumption that the amplitude
is a constant and integrate the cross section in \eqref{CS-pi} over angles, 
\begin{equation}
d\sigma =\frac{m_p^4 \pi^2}{(2\pi )^5 js}|{\cal M}_\pi|^2\, dM_{12}^2\,dM_{23}^2\;.
\end{equation}
For the total cross section we integrate over invariant masses
\begin{equation}
\sigma =\frac{m_p^{4} \pi^2}{(2\pi )^{5}js}|{\cal M}_\pi |^{2} I(s) \,,
\end{equation}
 where we introduced the integral
\begin{eqnarray}
I(s) &=& \int_{{M_{12}^{min}}^2}^{{M_{12}^{max}}^2}  \,
\Big( \int_{{M_{23}^{min}}^{2}}^{ {M_{23}^{max}}^{2}}
d M_{23}^2 \Big) \, dM_{12}^{2}  \nonumber \\
&=&  2 \int_{M_{12}^{min}}^{M_{12}^{max}} \sqrt{ (M_{12}^2 -4m_p^2 )} \nonumber \\
&& \times \sqrt{(s-m_\pi^2-M_{12}^2)^2-4 m_\pi^2 M_{12}^2}\, dM_{12}\,,
\eqlab{Is1}
\end{eqnarray}
with the lower and upper limits
$M_{12}^{min}=2m_p$ and $M_{12}^{max} = \sqrt{s}-m_\pi$ respectively. At the
upper limit the pion 3-momentum in the CM system,
$
\vec{k}^2 = {\left(M_{12}^2-(\sqrt{s} -m_\pi)^2 \right)  
\left(M_{12}^2-(\sqrt{s} +m_\pi)^2 \right)/ 4 s} \;,
$
vanishes while it reaches a maximum at the lower limit,
\beq
\vec{k}^2_{max}=m_\pi^2 \eta^2=b (b+4m_\pi \sqrt{s})/4s
\eqlab{eta}
\end{equation}
which defines the conventionally used variable $\eta$~\cite{Mey92} in terms of 
$b=(\sqrt{s} -m_\pi)^2-4m_p^2$. Introducing the relative $pp$ 3-momentum 
$
4\vec{p}^2=\left(M_{12}^2-4 m_p^2\right)
$
the phase space integral \eqref{Is1} can also be put in the familiar form
$ I(s) = \int 16 |\vec{p}| |\vec{k}| \sqrt{s / M_{12}^2} d\vec{p}^2
$.

With the substitutions $t=(M_{12}^2-4m_p^2)/b$, $a=4m_p^2$ and 
$c=(\sqrt{s} + m_\pi)^2-4m_p^2$ the phase space integral \eqref{Is1} can 
be cast in the form
\beq
I(s) = b^2 \int_0^1 dt \sqrt{ t(t-1)(t-c/b)\over(t+a/b) }\;.
\end{equation}
This shows that the total cross section is roughly proportional to 
$b^2$ and thus proportional to $\eta^4$ at low energies, in contrast to 
the data. It has been argued in \cite{Mey92} 
that the assumption made that the matrix element is constant is 
not valid. The pion-emission process is strongly affected by the final-state 
NN interaction at low energies, the effect of which is roughly 
proportional to $1/\vec{p}^2=4/t\,b$.
Including this factor in the integrand gives a total
cross section proportional to $\eta^2$, in rough agreement with
the data (above $\eta=0.2$). The approach that we chose for evaluating
the amplitude and comparison with experiment are described in detail
in sect. III.

\end{document}